# Buffer Requirements For TCP/IP Over ABR[1]

Shiv Kalyanaraman, Raj Jain, Sonia Fahmy, and Rohit Goyal
The Ohio State University
Department of CIS
Columbus, OH 43210-1277
Email: {*shivkuma, jain, goyal, fahmy*}@cis.ohio-state.edu

and Seong-Cheol Kim
Principal Engineer, Network Research Group
Communication Systems R&D Center
Samsung Electronics Co. Ltd.
Chung-Ang Newspaper Bldg.
8-2, Karak-Dong, Songpa-Ku
Seoul, Korea 138-160
Email: kimsc@metro.telecom.samsung.co.kr,

**Abstract**

We study the buffering requirements for zero cell loss for TCP over ABR. We show that the maximum buffers required at the switch is proportional to the maximum round trip time (RTT) of all VCs through the link. The number of round-trips depends upon the the switch algorithm used. With our ERICA [2] switch algorithm, we find that the buffering required is independent of the number of TCP sources. We substantiate our arguments with simulation results.

## 1 Introduction

ATM networks provide four classes of service: constant bit rate (CBR), variable bit rate (VBR), available bit rate (ABR), and unspecified bit rate (UBR). Data traffic in ATM is expected to be transported by the ABR service. The ATM Forum Traffic Management group has standardized a rate-based closed-loop flow control model for this class of traffic [5]. In this model, the ATM switches give feedback (explicit rate (ER) or binary (EFCI)) in Resource Management (RM) cells and the sources adjust their transmission rates appropriately. ATM switches use a scheme, like ERICA [2], to calculate the feedback. The details of ERICA are presented in reference [2] and we do not discuss it further in this paper.

TCP is the most popular transport protocol for data transfer. It provides a reliable transfer of data using a window-based flow and error control algorithm [6]. When TCP runs over ABR, the TCP window-based control runs on top of the ABR rate-based control. It is, hence, important to verify that the ABR control performs satisfactorily for TCP/IP traffic.

In a recent study [1], we examined the throughput and loss behavior of TCP over ABR with limited buffers. We observed a considerable drop in throughput even though the CLR was very small. In this paper, we quantify the buffer requirements for ABR to achieve the maximum TCP throughput with zero loss. We argue that the buffer requirement depends upon the round-trip time (RTT), the time delay for network feedback to take effect (feedback delay), the switch algorithm and its parameters, and the nature of higher priority background traffic, if any. The buffer requirement is independent of the number of TCP sources using the network.

In a separate study, we have shown that the UBR traffic class requires more buffers than ABR. The buffers required for TCP over UBR is proportional to the sum of all the TCP receiver windows [3].

---

[1] Proc. IEEE ATM'96 Workshop, San Fransisco, August 23-24, 1996.
Available from http://www.cis.ohio-state.edu/~jain/papers/atm96.ps

## 2  TCP And ERICA Options

We experiment with an infinite TCP source running on TCP over an ATM WAN. The TCP source always has a frame to send. However, due to TCP window constraint, the resulting traffic at the ATM layer may or may not be continuous. We use a TCP maximum segment size (MSS) of 512 bytes. The window scaling option is used so that the throughput is not limited by path length. The TCP window is set at 16x64 kB = 1024 kB. The zero-loss buffer requirement applies for all TCP congestion algorithms including "fast retransmit and recovery" algorithms. These algorithms are equivalent since there is no packet loss (assuming that spurious timeouts do not occur).

The ERICA algorithm uses two key parameters: target utilization and averaging interval length. The algorithm measures the load and number of active sources over successive averaging intervals and tries to achieve a link utilization equal to the target. The averaging intervals end either after the specified length or after a specified number of cells have been received, whichever happens first. In the simulations reported here, the target utilization is set at 90%, and the averaging interval length defaults to 100 ABR input cells or 1 ms, represented as the tuple (1ms,100 cells).

## 3  The $n$ Source Configuration

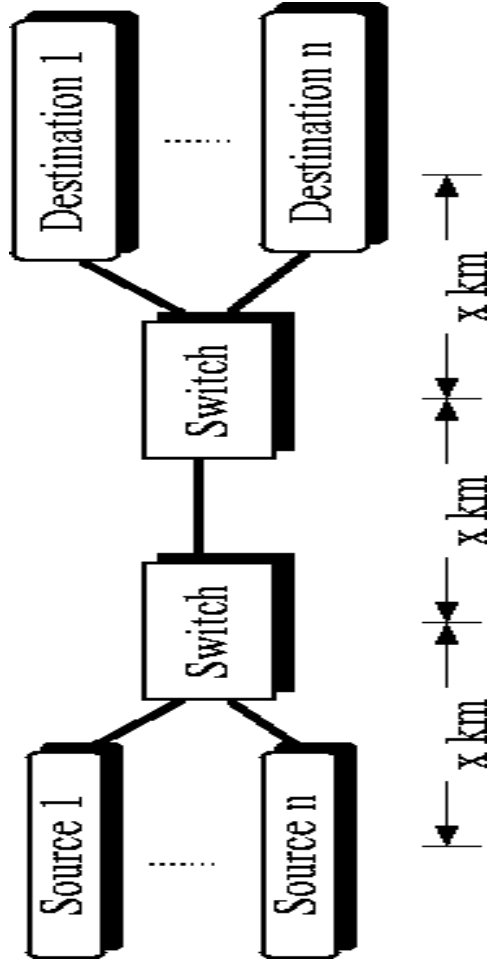

Figure 1: $n$ Source Configuration

The $n$ Source configuration has a single bottleneck link shared by the $n$ ABR sources. All links run at 155.52 Mbps and are of the same length. We experiment with the number of sources, the link lengths.

All traffic is unidirectional. A large (infinite) file transfer application runs on top of TCP for the TCP sources. N may assume values 1, 2, 5, 10, 15 and the link lengths 1000, 500, 200, 50km. We also summarize results with heterogenous link lengths.

Our performance metrics are: the *maximum ABR queue length* and *efficiency*. Efficiency is defined as the ratio of the sum of individual throughouts and the maximum possible throughput. The maximum possible throughput is calculated by accounting for the headers of a TCP segment over ATM. Assuming a maximum segment size of 512 bytes data, we have 20 bytes of TCP headers, 20 bytes of IP headers, 8 bytes LLC, 8 bytes AAL5 header. With the padding, this comes to 12 ATM cells = $12 \times 53 = 636$ bytes at the ATM layer. The maximum possible throughput is 512/636th of the link rate. On a 155.52 Mbps link, this works out to 125.2 Mbps. Note that ABR loses another 6% for RM cells (one out of every 32 cells) giving a maximum of 117.7 Mbps.

# 4 Summary Of Results

## 4.1 Observations About TCP Over ABR

ABR performance depends heavily upon the switch algorithm used. The following statements are based upon the ERICA switch algorithm [2]. In the following discussion, feedback delay from a particular switch (usually the bottleneck switch) refers to the time required by a cell to go from the switch to the source in the reverse direction and return to the switch from the source in the forward direction.

- There is no loss for TCP, if the switch has buffers at least (3×RTT+$c$×feedback delay)×link bandwidth, where $c$ is a constant which depends on the switch scheme and its parameters alone.

- The above buffering requirement is independent of the number of TCP sources. In other words, the same amount of buffers can sustain a very large number of ABR sources.

- Under many circumstances 1 × RTT × link bandwidth buffers may suffice.

- Drop policies improve throughput. But a proper drop policy is less critical than a proper switch algorithm.

- For ERICA, for a given target utilization, it appears that Qmax = ($a$×RTT+$b$×averaging interval length+ $c$ × Feedback delay) × link bandwidth provides a good approximation. In certain cases, the ERICA averaging interval length might be larger than the feedback delay, in which case, its effect dominates.

## 4.2 Informal Derivation (Summary)

The derivation of the (3×RTT+$c$×feedback delay)×link bandwidth requirement is based along the following arguments:

- Initially the TCP load doubles every RTT. During this phase, TCP sources are window-limited [1], i.e., their data transmission is bottlenecked by their congestion window sizes and not by the network directed rate.

- The minimum number of RTTs required to reach rate-limited operation [1] decreases as the logarithm of the number of sources. In other words, the more the number of sources, the faster they all reach rate-limited operation. Rate-limited operation occurs when the TCP sources are constrained by the network directed ABR rate rather than their congestion window sizes.

- After the pipe just becomes full (TCP keeps sending data for one RTT), the maximum queue which can build up before fresh feedback reaches the sources is 1 × RTT × link bandwidth. This observation follows because the aggregate TCP load can atmost double every RTT and fresh feedback reaches sources every RTT.

- Queue backlogs due to TCP bursts smaller than RTT (before the pipe became full) is 1 × RTT × link bandwidth. The TCP idle periods are not sufficient to drain out the queues built up during the TCP active periods. This occurs when the idle periods is shorter than the active periods. Given that TCP load doubles every RTT, the backlog is at most 1 × RTT × link bandwidth.

- Bursty behavior of ACKs causes an additional 1 × RTT × link bandwidth queues. When ACKs are bursty, the doubling of the TCP load can occur instantaneously (not spaced over time) and an extra round-trip worth of queues are built up.

- Once load is experienced continuously at the switch, the TCP sources appear as infinite sources to the switch. The switch algorithm then takes $c$ feedback delay times to converge to the max-min rates (when the queue length is guaranteed to decrease). Assuming that the TCP sources are rate-constrained during the convergence period, the aggegate TCP load can only decrease. In the worst case, the queue built up during the convergence phase is $c \times$ feedback delay $\times$ link bandwidth.

The sum of these components is approximately $(3 \times \text{RTT} + c \times \text{feedback delay}) \times$ link bandwidth.

## 5 Sample Simulation Results

In this section we present show sample simulation results to substantiate the preceding claims and analyses. The results presented here use ERICA without the fairness enhancement [2]. The queue lengths are slightly higher for ERICA with the fairness enhancement, due to the aggressive nature of the fairness algorithm.

### 5.1 Effect of number of sources

In Table 1, we notice that three RTTs worth of buffers are sufficient. One RTT worth of buffering is sufficient for many cases: for example, the cases where the number of sources is small. The rate distributions are fair in all cases.

Table 1: Effect of number of sources

| Number of Sources | RTT(ms) | Feedback delay(ms) | Max Q (cells) | Throughput | Efficiency |
|---:|---:|---:|---:|---:|---:|
| 5 | 30 | 10 | 10597 = 0.95*RTT | 104.89 | 83.78 |
| 10 | 30 | 10 | 14460 = 1.31*RTT | 105.84 | 84.54 |
| 15 | 30 | 10 | 15073 = 1.36*RTT | 107.13 | 85.57 |

### 5.2 Effect of Round Trip Time (RTT)

From table 2, we find that the maximum queues approaches of $3 \times \text{RTT} \times$ link bandwidth, particularly for MANs (6ms,1.5ms). This is because, the RTT values are lower and in such cases, the effect of switch parameters on the maximum queue increases. In particular, the ERICA averaging interval is comparable to the feedback delay.

Table 2: Effect of Round Trip Time (RTT)

| Number of Sources | RTT(ms) | Feedback Delay (ms) | Max Q size(cells) | Throughput | Efficiency |
|---:|---:|---:|---:|---:|---:|
| 15 | 30 | 10 | 15073 = 1.36*RTT | 107.13 | 85.57 |
| 15 | 15 | 5 | 12008 = 2.18*RTT | 108.00 | 86.26 |
| 15 | 6 | 2 | 6223 = 2.82*RTT | 109.99 | 87.85 |
| 15 | 1.5 | 0.5 | 1596 = 2.89*RTT | 110.56 | 88.31 |

### 5.3 LANs: Effect of Switch Parameters

In Table 3, the number of sources is 15. The averaging interval is the minimum of the time (T, in ms) and count (time for N input cells) values.

Table 3: Effect of Switch Parameter (Averaging Interval)

| Averaging Interval (ms,cells) | RTT(ms) | Feedback Delay (ms) | Max Q size(cells) | Thoughput | Efficiency |
|---|---|---|---|---|---|
| (10,500) | 1.5 | 0.5 | 2511 | 109.46 | 87.43 |
| (10,1000) | 1.5 | 0.5 | 2891 | 109.23 | 87.24 |
| (10,500) | 0.030 | 0.010 | 2253 | 109.34 | 87.33 |
| (10,1000) | 0.030 | 0.010 | 3597 | 109.81 | 87.71 |

From Table 3, we observe that, the effect of the switch parameters is pronounced in LAN configurations. The ERICA averaging interval becomes much greater than the RTT and feedback delay and determines the rate of feedback to the sources.

### 5.4 Effect of Feedback Delay

We conducted a 3 × 3 full factorial experimental design to understand the effect of RTT and feedback delays [4]. The results are summarized in Table 4. The thoughput and efficiency figures for the last three rows (550 ms RTT) are not available since the throughput did not reach a steady state although the queues had stabilized.

Table 4: Effect of Feedback Delay

| RTT(ms) | Feedback Delay (ms) | Max Q size(cells) | Thoughput | Efficiency |
|---|---|---|---|---|
| 15 | 0.01 | 709 | 113.37 | 90.55 |
| 15 | 1 | 3193 | 112.87 | 90.15 |
| 15 | 10 | 17833 | 109.86 | 87.75 |
| 30 | 0.01 | 719 | 105.94 | 84.62 |
| 30 | 1 | 2928 | 106.9 | 85.39 |
| 30 | 10 | 15073 | 107.13 | 85.57 |
| 550 | 0.01 | 2059 | NA | NA |
| 550 | 1 | 15307 | NA | NA |
| 550 | 10 | 17309 | NA | NA |

Observe that the queues are small when the feedback delay is small and do not increase substantially with round-trip time. This is because the switch scheme limits the rate of the sources before they can overload for a substantial duration of time.

## 6 Conclusions

In this study, we have observed that the ABR service is scalable in terms of number of sources (or virtual circuits). The total buffer required in a switch to achive zero loss is bounded. This bound depends upon the RTT of VCs but not on their number. Thus, a switch with buffers equal to a small multiple of network diameter can guarantee no loss even for a very large number of VCs. Of course, the applicability of this statement and the multiplication factor depend upon the switch algorithm, which has not been standardized. We have shown an existance proof using our ERICA switch algorithm.

Other factors that affect the bound are feedback delay (sum of delay from the bottleneck back to source and from source to the bottleneck) If the feedback delay is small and the switch scheme converges quickly, the effect of large round-trip delays can be reduced.

We have studied cases where the feedback delays are heterogenous and found that the queue is bounded by the Qmax calculated with the largest RTT and feedback delay. We also observed that as the number of sources with larger RTTs and feedback delays increased, the queues increased (though limited by the same bound). The results are omitted for lack of space.

ERICA algorithm has an additional option that allows the queue bound to be limited further. This option called queue-control or ERICA+ allows network managers to set a queue threshold goal and the algorithm tries to achive that goal. This option is particularly helpful if both the RTT and feedback delays are large [2].

In this paper, we have not studied the impact of higher priority background traffic such as CBR or VBR. In particular, presence of VBR introduces variance in load as well as capacity. When the variance is high, it is better to use the queue control feature of ERICA (known as ERICA+) to achieve the three goals of high utilization, fairness and bounded small queues. The effect of TCP and VBR will be the subject of a future paper.

---

[2] All our papers and ATM Forum contributions are available through http://www.cis.ohio-state.edu/~jain
[3] Throughout this section, AF-TM refers to ATM Forum Traffic Management sub-working group contributions.